\documentclass[11pt,reqno]{article}

\usepackage[english]{babel}
\usepackage{amsmath}
\usepackage{amssymb}
\usepackage{mathrsfs}
\usepackage{amsthm}
\usepackage{graphicx}
\usepackage{subfigure}

\def\p{\par\noindent}
\def\RE{\mathbb R}
\def\CO{\mathbb C}
\def\g{\mathcal G_\zeta}
\def\sgn{\text{\rm sgn}}
\def\l{\ell}
\def\S{\mathcal S}

\newcommand{\ve}{\varepsilon}
\newcommand{\be}{\begin{equation}}
\newcommand{\ee}{\end{equation}}
\newcommand{\n}{\noindent}
\newcommand{\HH}{\mathcal{H}}
\newcommand{\tHH}{\tilde{\mathcal{H}}}
\newcommand{\W}{\tilde{\mathscr{W}}}
\newcommand{\WW}{\tilde{W}}

\newcommand{\Z}{\mathbb{Z}}
\newcommand{\N}{\mathbb{N}}
\newcommand{\te}{\theta}
\newcommand{\ue}{\underline{e}}
\newcommand{\uz}{\underline{z}}
\newcommand{\uy}{\underline{y}}
\newcommand{\uphi}{\underline{\phi}}
\newcommand{\upsi}{\underline{\psi}}
\newcommand{\uchi}{\underline{\chi}}
\newcommand{\usigma}{\underline{\sigma}}

\newcommand{\tp}{\tilde p}
\newcommand{\tv}{\tilde v}
\newcommand{\ty}{\tilde y}
\newcommand{\tz}{\tilde z}
\newcommand{\tsig}{\tilde\sigma}
\newcommand{\pd}[2]{\frac{\partial {#1}}{\partial {#2}}}

\newtheorem{theorem}{Theorem}[section]
\newtheorem{lemma}[theorem]{Lemma}
\newtheorem{corollary}[theorem]{Corollary}
\theoremstyle{definition}

\theoremstyle{remark}

\numberwithin{equation}{section}

\begin{document}

\begin{center}
{\Large
{\bf Point Interactions in Acoustics: One Dimensional\\ Models.}\\
\bigskip
 C. Cacciapuoti\footnote{claudio.cacciapuoti@na.infn.it}, R. Figari\footnote{figari@na.infn.it},
A. Posilicano\footnote{andrea.posilicano@uninsubria.it}\\
\vspace{1cm}
$^{1,2}$Istituto Nazionale di Fisica Nucleare, Sezione di Napoli\\
Dipartimento di Scienze Fisiche\\
Universit\`a di Napoli Federico II,\\
Via Cintia 80126 Napoli, Italy\\
\vspace{0,5cm}
$^3$Dipartimento di Fisica e Matematica\\
Universit\`a dell'Insubria\\
Via Valleggio 11,  22100 Como, Italy\\}
\end{center}

\vspace{1cm}

\begin{abstract}
A one dimensional system made up of a compressible fluid and several
mechanical oscillators, coupled to the acoustic field in the fluid,
is analyzed for different settings of the oscillators array. The
dynamical models are formulated in terms of singular perturbations of
the decoupled dynamics of the acoustic field and the mechanical
oscillators.

\n
Detailed spectral properties  of the generators of the dynamics are
given for each model we consider. In the case of a periodic array of
mechanical oscillators it is shown that the energy spectrum presents
a band structure.
\end{abstract}

\newpage

\section*{Introduction}

\n In this paper we study the dynamics of a system consisting of one or more
mechanical oscillators (the sources) coupled with the acoustic field
they produce in the compressible fluid surrounding them.

\n
Classical electromagnetism is perhaps the most well known case in theoretical
physics where all attempts to construct a complete, covariant, causal,
divergence free
theory for the evolution of the fields together with their sources
were unsuccessful up to now (in fact it is hard
to say that there is a single case in Classical or in Quantum Physics in which
this problem was completely solved).

\n Whereas theories with extended rigid charges are quite well
understood both
at the classical and the quantum level (see
e.g. the recent book \cite{Spohn} for a systematic introduction to the
subject and for a long list of references), there is no mathematically
consistent theory of point charges interacting with their own electromagnetic field. Indeed Newton equations with
Lorentz force require the fields to
be evaluated at the particle positions, and this produces infinities
due to the presence of the point-like sources.
These difficulties directly lead to the
need of mass renormalization. In his seminal paper Dirac \cite{D}
(also see \cite{IW}, \cite{K}, \cite{M}),
without using Lorentz force but exploiting the conservation of
energy and momentum and considering their flow through a
thin tube of radius $r$, derived an equation for the motion of
a
charged point particle (the Lorentz-Dirac equation). As Dirac himself
pointed out the equation obtained in the limit $r\downarrow 0$,
together with the mass renormalization, leads to the presence of
runaway solutions, i.e. solutions for which the
acceleration increases beyond any bound even in the absence of
external fields.

\n  An approach
based on the theory of singular perturbations of the free dynamics was initiated in \cite{NP1} and \cite{NP2}
for the case of classical
electrodynamics of a point particle in the dipole (or linearized)
case. Here the generator of the limit dynamics
of both the field and the particle appears to be a singular
perturbation of the generator of the free dynamics. The
phenomenological mass plays the role of the parameter describing a suitable
family of self-adjoint extensions and the boundary
condition naturally appearing in the domain of the generator results
to be nothing else that a regularized
(and linearized) version of the usual velocity-momentum relation in
the presence of an electromagnetic field.
In this framework runaway
solutions are unavoidable because a negative
eigenvalue appears in the spectrum of the generator after mass
renormalization.

\n Our interest in a similar problem in acoustics was prompted by the appearance in 1999
of a paper by J. D. Templin \cite{T}. In that paper the author analyzed the dynamics of a
simple model of a spherical oscillator interacting with the acoustic field it
generates. The existence of a spherically symmetric
radiation field (the \emph{acoustic monopole}) makes the acoustic
case significantly different from the electromagnetic one. Moreover
the pressure field at the surface of the sphere completely
characterizes the contact forces responsible of the interaction
between source and field in the acoustic case.

\n Templin performed a detailed analysis of the field emitted by the
acoustic monopole, explicitly computing both its radiation and
near-field components. He noticed that a deduction of the
\emph{reaction field} obtained from the emitted radiation power,
therefore neglecting
the near field component, brings to an equation for the radius of the
oscillating sphere showing \emph{runaway solutions}.

\n In analogy with what was done for the electromagnetic case in
\cite{NP1} we want to provide a formalization of the problem of a
finite or infinite number of oscillators coupled with their acoustic
field in terms of singular perturbations of the generator of the free
dynamics.

\n
In this paper we will consider only the one dimensional case. In
the abstract setting we will work in the
physical model of interaction between sources and field will appear as the only possible extension of the free dynamics. The generalization to three dimensions is not straightforward.
From one side a  model of a physically relevant, symmetric, mechanical oscillator
with finite degrees of freedom is lacking. On the other side point
perturbations of the free dynamics are much more singular in higher
dimensions. We plan to discuss the three dimensional case in further
work.

\n
We want to stress an aspect of the dynamical system we analyze here which was extensively studied in different contexts.
As an immediate consequence of the third Newton's law and of the assumption
of persistent contact between the fluid and the surface of the oscillators, the total energy, sum of  the (positive) energy of the
acoustic field $E_{ac}$ and the (positive)  energy of the oscillators $E_{osc}$,
is a constant of motion. As an immediate consequence one can exclude the existence of runaway solutions in this case. Moreover, lacking a mechanism of reflection
of the acoustic waves at some exterior boundary, the motion of the oscillators should be damped
and the energy should finally diffuse over the field degrees of freedom,
for almost every initial condition. The situation
is reminiscent of the one investigated in \cite{SW1}, \cite{SW2} and
\cite{SW3} about the diffusion of energy from bound
states to continuous states triggered by time dependent perturbations
in quantum and classical systems. In our system there is no external potential the interaction being given by internal forces.

\n
This paper is organized as follows.

\n
In section \ref{secone} we introduce a list of
notation and we briefly recall the equations for the acoustic field.
Afterwards we exemplify the problem of the interaction between the
field and a source in the completely solvable case of a single wall
attracted toward the origin by a linear restoring force.

\n
In section \ref{sectwo} we analyze the case of a finite number of
sources in the framework of the possible extensions of the free
dynamics outside the points where the sources are placed.

\n
In section \ref{secthree} we generalize the construction to the case of infinitely many sources and study the
case of sources periodically placed on the real line. We give detailed results on the
characteristic band
structure of the spectrum of the generator of the dynamics.

\n To the best of our knowledge this kind of systems of oscillators coupled with the acoustic field was never proposed and solved. A remark on the band structure of a similar model is in \cite{GS}.

\section{\label{secone}The acoustic monopole in one dimension}

\n We give a detailed description of our model in the simplest case of
one oscillator coupled with the acoustic field.

\n Consider an infinite pipe filled with a non viscous, compressible fluid. We suppose that there is no friction between the fluid and the pipe and we choose a coordinate system with the $x$-axis parallel to the axis of the pipe. The mechanical
oscillator is made up of a very thin wall of mass $M$ positioned in the pipe
perpendicularly to the axis in $x=0$. The thin wall is connected to a spring of elastic constant $K$. We analyze only one dimensional cases, hence the acoustic field is described by the pressure field  $p(x,t)$ and the velocity field $v(x,t)$. The motion of the mechanical oscillator is described through the position and the velocity of the thin wall.

\n The field $p(x,t)$ represents deviations of the pressure in the point $x$ at time $t$ with respect
to an equilibrium pressure $P_0$. In the linearized
acoustics regime the continuity equation, the Newton's second law and the adiabatic equation of state read
\be
\pd{\rho}{t}+\rho_0\pd{v}{x}=0\;,\quad
\rho_0\pd{v}{t}=-\pd{p}{x}\;,\quad
p=a^2\rho\;,
\ee
\n where $\rho(x,t)$ is the deviation  of the density in the point $x$ at time
 $t$ with respect to the equilibrium density $\rho_0$ and $a$ is the velocity of sound in the fluid.

\n Then we have for $p(x,t)$ and $v(x,t)$ the following coupled
differential equations
\be
\pd{p}{t}=-a^2\rho_0\pd{v}{x}\;,\quad
\pd{v}{t}=-\frac{1}{\rho_0}\pd{p}{x}\;.
\label{pveqs}
\ee

\n We consider only small oscillations of the thin wall around
  its equilibrium position $x=0$, we indicate with $y(t)$ the displacement of the wall
from its equilibrium position at time $t$ and we suppose that
the wall remains always in contact with the fluid
\be
v(y(t),t)=\frac{dy(t)}{dt}\qquad\forall t\geq0\;.
\label{vyteqyt}
\ee
\n Notice that we consider a wall of zero thickness. We make
the approximation $v(y(t),t)\simeq v(0,t)$ and  condition
(\ref{vyteqyt}) becomes
\be
v(0,t)=\frac{dy(t)}{dt}\qquad\forall t\geq0\;.
\label{v0teqyt}
\ee

\n The equation of motion for the position of the thin wall $y(t)$ is
\be
M\ddot y(t)=-Ky(t)-S\left(p(0^+,t)-p(0^-,t)\right)
\label{eqosc}
\ee
\n where $S$ is the area of the transverse section of the pipe and we made the
approximation $p(y^\pm(t),t)\simeq p(0^\pm,t)$.

\n The total energy of the system is given by
\be
E_{tot}=E_{ac}+E_{osc}
\ee
\n with
\begin{align}
E_{ac}&=\frac{S}{2a^2\rho_0}\int_{-\infty}^\infty
p(x)^2dx+\frac{S\rho_0}{2}\int_{-\infty}^\infty
v(x)^2dx\\
E_{osc}&=\frac{K}{2}y^2+\frac{M}{2}\dot y^2\,,
\end{align}
\n $E_{ac}$ is the energy stored in the acoustic field while $E_{osc}$
is the energy of the mechanical oscillator.

\n As the system is isolated the energy is constant. The motion of the wall
produces acoustic waves thus transferring continuously energy from the
oscillator to the
acoustic field. One then expects that $y(t)$ decreases to zero when $t\to\infty$.

\n In spite of being a simple exercise, the exact computation of the solution of problem (\ref{pveqs}), (\ref{v0teqyt}), (\ref{eqosc}) and, in turn, of the damping rate of the oscillations rarely appears in textbooks.

\n In the following we give the solution of the Cauchy problem of
coupled ordinary and partial differential equations with time dependent
boundary conditions
\be
\begin{cases}
\displaystyle\pd{p}{t}=-a^2\rho_0\pd{v}{x}&\forall t\geq0\;\forall x\in\RE\backslash\{0\}\\
\displaystyle\pd{v}{t}=-\frac{1}{\rho_0}\pd{p}{x}&\forall t\geq0\;\forall x\in\RE\backslash\{0\}\\
\ddot{y}(t)=-\omega_0^2
y(t)-\displaystyle\frac{S}{M}\left(p(0^+,t)-p(0^-,t)\right)&\forall t\geq0\\
p(x,0)=f(x)&\forall x\in\RE\backslash\{0\}\\
v(x,0)=g(x)&\forall x\in\RE\backslash\{0\}\\
y(0)=y_0&\\
\dot y(0)=\dot y_0&\\
v(0,t)=\dot y(t)&\forall t\geq0
\label{directproblem}
\end{cases}
\ee
\n where $f(x)$ and $g(x)$ are two real functions and $\omega_0^2=K/M$.

\n Suppose that
\be
f(x)\in C_0^2(\RE)\,;\; g(x)\in C_0^2(\RE)\quad\textrm{and}\quad
y_0=\displaystyle\frac{1}{\omega_0^2\rho_0}f'(0)\,;\;\dot y_0=g(0)\,,
\label{regcond}
\ee
\n then the solution of problem (\ref{directproblem}) reads
\begin{align}
p(x,t)&=p_f(x,t)+a\rho_0\;\sgn(x) Y\left(t-\frac{|x|}{a}\right)\\
v(x,t)&=v_f(x,t)+Y\left(t-\frac{|x|}{a}\right)\\
y(t)&=-\frac{\dot{v}_f(0,t)}{\omega_0^2}+\frac{\displaystyle\int_0^t\frac{F(t')}{\beta_+}e^{\beta_+(t-t')}dt'-
\int_0^t\frac{F(t')}{\beta_-}e^{\beta_-(t-t')}dt'}{\beta_+-\beta_{-}}\label{yt}
\end{align}
\n where $p_f(x,t)$ and $v_f(x,t)$ are solution of the wave equation in $(-\infty,+\infty)$ with initial conditions $f(x)$ and $g(x)$
\be
\begin{aligned}
p_f(x,t)&=\frac{f(x-at)+f(x+at)}{2}+
\frac{a\rho_0}{2}\left(g(x-at)-g(x+at)\right)\\
v_f(x,t)&=\frac{g(x-at)+g(x+at)}{2}+
\frac{1}{2a\rho_0}\left(f(x-at)-f(x+at)\right)\,,
\end{aligned}
\ee
\be
F(t)=-\ddot{v}_f(0,t)-\omega_0^2v_f(0,t)
\ee
\n and
\be
Y(t)=\frac{\displaystyle\int_0^tF(t')e^{\beta_+(t-t')}dt'-
\int_0^tF(t')e^{\beta_-(t-t')}dt'}{\beta_+-\beta_{-}}
\ee
\n with $\beta_{\pm}=(-\gamma\pm\sqrt{\gamma^2-4\omega_0^2})/2$ with
$\gamma=2a\rho_0 S/M$.

\n By conditions (\ref{regcond}) one
easily obtain that $y(t)$ and $\dot y(t)$ are both continuous and decrease exponentially to
zero with decay constant $\tau=\gamma/2$.

\section{\label{sectwo}Singular perturbations of the free dynamics\label{ssingular}}

\n   In  this  section   we  present   a  generalization   of  problem
(\ref{directproblem}) formulated in terms of a unitary flow on a space
of finite energy.

\n Let us consider a system of $n$ thin walls positioned in a pipe, perpendicularly
to its axis. Let
$\S=\left\{s_1,\dots,s_n\right\}\subset\RE$ be the set of
equilibrium positions of the thin walls. The $i$-th thin wall,
placed in $s_i$, has mass  $M_i$ and is connected to a spring of
elastic constant $K_i$. The acoustic field is described by the
pressure field  $p$ and the velocity field $v$. The motion of the walls
is described by the displacements $y_j$ from their equilibrium positions
and by the corresponding velocities $z_j$.

\n The generator of the dynamics, $\hat A$, will be defined as a singular
perturbation of the skew-adjoint operator $A$ generating the uncoupled
evolution of the acoustic field and of the oscillators.

\n The system of first order differential equations
\begin{align}
\frac{\partial p}{\partial t}&=-a^2\rho_0\,\frac{\partial v}{\partial
x}\label{dueuno}&\forall x\in\RE\\
\frac{\partial v}{\partial t}&=-\frac{1}{\rho_0}\,\frac{\partial
p}{\partial x}&\forall x\in\RE\\
\frac{dy_j}{dt}&=z_j&1\leq j\leq n\\
\frac{dz_j}{dt}&=-\frac{K_j}{M_j}\,y_j&1\leq j\leq
n\label{duequattro}
\end{align}

\n describes, in the linear approximation, the independent  evolution  of  $n$  mechanical
oscillators  and  of  the acoustic field.

\n We want to  show first how equations  (\ref{dueuno}) -
(\ref{duequattro}) define an unitary flow in a complex Hilbert
space. To this aim let us consider equations  (\ref{dueuno}) -
(\ref{duequattro}) for complex functions  $v$, $p$, $y_j$, $z_j$ of
position and time.

\n The set of all the displacements and velocities will be represented respectively  by the
vectors in $\CO^n$\be \uy=y_1\ue_1+\dots+y_n\ue_n\,,\quad
\uz=z_1\ue_1+\dots+ z_n\ue_n \,, \ee where $\ue_1,\dots,\ue_n$ is
the canonical orthonormal basis in $\CO^n$.

\n Let us denote by $L^2(\RE)$ the space of square-integrable functions on
the real line. $\bar H^1(\RE)$ indicates the homogeneous Sobolev space of
locally square-integrable functions with square-integrable (distributional)
derivative, and $H^1(\RE)$ the usual Sobolev space $H^1(\RE)
:=\bar H^1(\RE)\cap
L^2(\RE)$.

\n Therefore the linear operator $A$ in
$L^2(\RE)\oplus L^2(\RE)\oplus\CO^n\oplus\CO^n$ generating the
dynamics (\ref{dueuno}) - (\ref{duequattro}) is defined by
\be
A: H^1(\RE)\oplus H^1(\RE)\oplus\CO^n\oplus\CO^n
\to L^2(\RE)\oplus L^2(\RE)\oplus\CO^n\oplus\CO^n
\ee
\be
A(p,v,\uy,\uz):=\left(-a^2\rho_0\,\frac{dv}{dx},\,-\frac{1}{\rho_0}\,
\frac{dp}{dx},\,\uz,\,-\sum_{1\le j\le n}\frac{K_j}{M_j}\, y_j\ue_j\right)\,,
\ee
\n where $a$, $\rho_0$, $K_j$,  $M_j$, $1\le j\le n$,
are the positive real constants representing the physical parameters.

\n In the following a capital Greek letter will indicate a generic vector $(p,v,\uy,\uz)$ in
$L^2(\RE)\oplus L^2(\RE)\oplus\CO^n\oplus\CO^n$.

\n $A$ is a real operator, i.e.
it preserves the (physical) linear subspace of real elements
\be\label{phys}
\left\{(p,v,\uy,\uz)\,:\, p(x)\in\RE,\ v(x)\in\RE,\ \uy\in\RE^n,\
\uz\in\RE^n\right\}\,.
\ee
$A$ is skew-symmetric with respect to the
scalar product
\be
\langle\langle\Psi_1,\Psi_2\rangle\rangle\equiv\frac{1}{a^2\rho_0}\,\langle p_1,p_2\rangle+\rho_0
\langle v_1,v_2\rangle+\frac{1}{S}\sum_{1\le j\le n}{K_j}\, \bar y_{1j}
y_{2j}
+{M_j}\,\bar z_{1j}z_{2j}\,,
\label{scalprod}
\ee
where $\langle\cdot\,,\cdot\rangle$ indicates the standard scalar product in $L^2(\RE)$, $S$ is the area of the transverse section of the pipe and $^-$ denotes
complex conjugation. $L^2(\RE)\oplus L^2(\RE)\oplus\CO^n\oplus\CO^n$ is a Hilbert
space with the scalar product (\ref{scalprod}).

\n The square norm of a vector $\Psi$, $\|\Psi\|^2=\langle\langle\Psi,\Psi\rangle\rangle$,
defines the total energy of the system in the state $\Psi$
\be
E_{tot}=\frac{S}{2}\|\Psi\|^2=E_{ac}+E_{osc}
\ee
\n where $E_{ac}$ is the energy stored in the acoustic field while
$E_{osc}$ is the energy of the oscillators
\be
E_{ac}=\frac{S}{2a^2\rho_0}\,\langle p,p\rangle+\frac{\rho_0 S}{2}
\langle v,v\rangle\;;\quad
E_{osc}=\frac{1}{2}\sum_{1\le j\le n}\left({K_j} |y_{j}|^2
+{M_j}\, |z_{j}|^2\right)\,.
\ee

\n For any $\zeta\in\CO\backslash i\RE$ the resolvent of $A$ is
\be
\begin{aligned}
&(-A+\zeta)^{-1}(p,v,\uy,\uz)=\left(
\rho_0\left(-\frac{d^2}{dx^2}+\frac{\zeta^2}{a^2}\right)^{-1}
\left(-\frac{dv}{dx}+\frac{\zeta}{a^2\rho_0}\,p\right),\right.\\
&\frac{1}{a^2\rho_0}\,\left(-\frac{d^2}{dx^2}
+\frac{\zeta^2}{a^2}\right)^{-1}
\left(-\frac{dp}{dx}+\zeta\rho_0 v\right),\
\sum_{1\le j\le n}\frac{M_jz_j+\zeta M_jy_j}{K_j+\zeta^2M_j}\,\ue_j\,, \\
&\left.\sum_{1\le j\le n}\frac{-K_jy_j+\zeta
M_jz_j}{K_j+\zeta^2M_j}\,\ue_j
\right)\,.
\end{aligned}
\ee Since \be \text{\rm Ran$(-A\pm 1)$}=L^2(\RE)\oplus
L^2(\RE)\oplus\CO^n\oplus\CO^n\,, \ee $A$ is skew-adjoint. Moreover
the essential spectrum of $A$ is purely absolutely continuous and
\be \sigma_{ess}(A)=\sigma_{ac}(A)=i\RE\,,\quad
\sigma_{pp}(A)=\left\{\pm i\,\sqrt{\frac{K_j}{M_j}}\,,\ 1\le j\le
n\right\}\,. \ee Being skew-adjoint the operator $A$ describes, by
Stone theorem, the uncoupled evolution of the acoustic field and of
the oscillators through the unitary flow $\exp t A$ corresponding to
the Cauchy problem for the first order differential equation \be
\frac{d}{dt}\Psi(t)=A\Psi(t) \ee which is equivalent to the system
written at the beginning of the section.\p Now we consider the
linear operator $A_0$ obtained by restricting $A$ on the set of
vectors in its domain satisfying: \be \left\{v(s_j)=z_j\,,\quad 1\le
j\le n\right\} \ee which represents the kinematic  constraint
(\ref{v0teqyt}) at each thin wall. $A_0$ is a closed, densely
defined, skew-symmetric linear operator with defect indices $(n,n)$.
We want to characterize the skew-adjoint extensions of $A_0$. The
family of extensions of $A_0$ can be parameterized by relations
$K\subset\CO^n\oplus \CO^n$ which are skew-symmetric, i.e. such that
$K=({\mathscr I}K)^\perp$, where ${\mathscr
I}(\uz_1,\uz_2):=(\uz_2,\uz_1)$ (see e.g. \cite{GG}, Theorem 1.6,
chapter 3, for the analogous self-adjoint case). A skew-symmetric
relation in $\CO^n\oplus \CO^n$ extends the notion of the graph of a
skew-symmetric operator $\Theta:\CO^n\to\CO^n$ through the relation
$K=\{(\uz,\Theta\uz) \,,\  \uz\in\CO^n\}$. In  order  to  be a
candidate  to describe  the interacting dynamics of the system under
analysis a skew-adjoint extension of  $A_0$ must be local and real
i.e. it must generate a coupling between the fields evaluated in
$s_j$ and the $j$-th oscillator, $1\leq j\leq n$, and it must
preserve the linear space of physical data defined in (\ref{phys}).
The only admissible extension different from $A$ itself will be the
one corresponding to the graph of the zero operator, $\Theta=0$. The
next theorem completely characterizes such an extension:
\begin{theorem}
The only local, real and skew-adjoint extension of $A_0$ is given by
\be \hat A:D(\hat A)\subset L^2(\RE)\oplus
L^2(\RE)\oplus\CO^n\oplus\CO^n\to L^2(\RE)\oplus
L^2(\RE)\oplus\CO^n\oplus\CO^n\,, \ee \be
\begin{aligned}\label{dom}
D(\hat A)
=\{&\Psi\equiv(p,v,\uy,\uz)\,
:\, p\in L^2(\RE)\cap H^1(\RE\backslash \S),\, v\in
H^1(\RE),\  \uy\in\CO^n,\\ & \uz\in\CO^n,\ p(s_i^+)-p(s_i^-)=\sigma_i,\
v(s_j)=z_j,\ \usigma\in \CO^n\}\,,
\end{aligned}
\ee
\be
\begin{aligned}
&\hat A(p,v,\uy,\uz):=
\\&:=\left(-a^2\rho_0\,\frac{dv}{dx},\,
-\frac{1}{\rho_0}\,\frac{dp_0}{dx},\,\uz,\, -\sum_{1\le j\le
n}\left(\frac{K_j}{M_j}\,y_j+\frac{S}{M_j}\,\sigma_j\right)\,\ue_j\right)\,.
\end{aligned}
\ee
Here $p_0\in \bar H^1(\RE)$,
\be
p_0(x):=p(x)-\frac{1}{2}\sum_{1\le j\le
n}\sigma_j\,
\sgn(x-s_j)\,,
\ee
denotes the regular part of $p$. The resolvent of $\hat A$ is given by
\be
(-\hat A+\zeta)^{-1}=
(-A+\zeta)^{-1}+\sum_{1\le i,j\le n}\left(\Gamma(\zeta)^{-1}\right)_{ij}\,
G^i_\zeta\otimes
\breve G^j_{\bar\zeta}\,,\qquad \zeta\in\CO\backslash i\RE\,,
\ee
where
\be
\Gamma(\zeta)_{ij}
:=
-\zeta\left(\pm\frac{e^{\mp\zeta|s_i-s_j|/a}}{2a\rho_0\zeta}
+\frac{S\delta_{ij}}{K_j+\zeta^2M_j}\right)\,,\quad\text{\rm $\pm$Re$\,\zeta>0$}
\label{gamma}
\ee
and
\be
\breve G_\zeta^j(x)=\left(
\g'(x-s_j)\,, \frac{\zeta}{a^2\rho_0}\,\g(x-s_j)\,,
\frac{S}{K_j+\zeta^2M_j}\,\ue_j\,,\
\frac{-\zeta S}{K_j+\zeta^2M_j}\,\ue_j
\right)\,,
\ee
\be
G_\zeta^j(x)=\left(
-\g'(x-s_j)\,, \frac{\zeta}{a^2\rho_0}\,\g(x-s_j)\,,
\frac{-S}{K_j+\zeta^2M_j}\,\ue_j\,,\
\frac{-\zeta S}{K_j+\zeta^2M_j}\,\ue_j
\right)\,,
\ee
\be
\g(x)=\pm \frac{a}{2\zeta}\,e^{\mp\zeta|x|/a}\,,\quad
\g'(x)=-\frac{1}{2}\,\sgn(x)\,e^{\mp\zeta|x|/a}\,,\quad \text{\rm $\pm$Re$\,\zeta>0$}\,.
\ee
\end{theorem}
\begin{proof} Since $iA_0$ is a closed, densely defined, symmetric
operator with defect indices $(n,n)$, all its self-adjoint
extensions can be obtained by the famed von Neumann theory on
self-adjoint extensions (see e.g. Theorem X.2 in \cite{RS2}).
However, since $A_0$ is obtained by restricting the skew-adjoint
operator $A$ to the kernel of the continuous, surjective linear
operator \be \tau: H^1(\RE)\oplus
H^1(\RE)\oplus\CO^n\oplus\CO^n\to\CO^n\,, \ee \be
\tau(p,v,\uy,\uz):=\sum_{1\le j\le n}(v(s_j)-z_j)\,\ue_j\,, \ee it
is easier to make use of the (equivalent) procedure developed in
\cite{P1} (also see the appendix in \cite{P2} for a compact review).
Here below we provide the (almost) self-contained construction of
the skew-adjoint extensions of $A_0$ by using such a procedure. \p
Given the map $\tau$ we can define the bounded linear operator \be
\breve G(\zeta):=\tau(-A+\zeta)^{-1}: L^2(\RE)\oplus
L^2(\RE)\oplus\CO^n\oplus\CO^n\to\CO^n\,. \ee \n By the relation \be
\breve G(\zeta)(p,v,\uy,\uz)=\sum_{1\le j\le n} \langle\langle
\breve G^j_{\bar\zeta},(p,v,\uy,\uz)\rangle\rangle\,\ue_j\,,\quad
1\le j\le n\,, \ee $\breve G(\zeta)$ is represented by the vector
$\breve G_\zeta^j$. By $\breve G(\zeta)$ we define the bounded
linear operator \be G(\zeta):=-\,\breve G(-\bar\zeta\,)^*:\CO^n\to
L^2(\RE)\oplus L^2(\RE)\oplus\CO^n\oplus\CO^n\,, \ee \n where
$\breve G(\zeta)^*$ indicates the adjoint of $\breve G(\zeta)$. The
action of $G(\zeta)$ on $\CO^n$ is given by \be
G(\zeta)\,\ue_j=G^j_\zeta\,,\qquad 1\le j\le n\,. \ee Let us notice
that \be \text{\rm Ran$(G(\zeta))$}\cap H^1(\RE)\oplus
H^1(\RE)\oplus\CO^n\oplus\CO^n =\left\{0\right\}\,. \label{rangzeta}
\ee Now we consider the linear operator
$\Gamma_\Theta(\zeta):\CO^n\to\CO^n$ represented by the matrix
$\Theta_{ij}+\Gamma(\zeta)_{ij}$, where $\Theta:\CO^n\to\CO^n$ is
skew-Hermitian. By noticing that \be
\Gamma_\Theta(\zeta)-\Gamma_\Theta(\xi)=\tau(G(\xi)-G(\zeta)) \ee
and that, by the definition of $G(\zeta)$ and by the first resolvent
identity, \be (\zeta-\xi)(-A+\xi)^{-1}G(\zeta)=G(\xi)-G(\zeta)\,,
\ee one has that $\Gamma_\Theta(\zeta)$ satisfies the identity \be
\Gamma_\Theta(\zeta)-\Gamma_\Theta(\xi)=(\zeta-\xi)\breve
G(\xi)G(\zeta)\,. \label{gammazetagammaxi} \ee By the definitions of
$\breve G(\zeta)$ and $G(\zeta)$, by (\ref{gammazetagammaxi}) and by
$\Gamma_\Theta(\bar\zeta)^*= -\Gamma_\Theta(-\zeta)$, it follows
that det$\,\Gamma_\Theta(\zeta)\not=0$ for any
$\zeta\in\CO\backslash i\RE$ and that \be \hat R(\zeta):=
(-A+\zeta)^{-1}+G(\zeta)\Gamma_\Theta(\zeta)^{-1}\breve G(\zeta) \ee
satisfies the first resolvent identity \be (\zeta-\xi)\,\hat
R(\xi)\hat R(\zeta)= \hat R(\xi)-\hat R(\zeta) \label{zeta-xi} \ee
and \be \hat R(\bar\zeta)^*=-\hat R(-\zeta) \label{hatR} \ee (for
details see \cite{P1}). \n Moreover $\hat R(\zeta)$ is injective by
(\ref{rangzeta}). Therefore \be \hat A:=-\hat R(\zeta)^{-1}+\zeta
\ee is well defined on \be\label{ran} D(\hat A):=\text{Ran$(\hat
R(\zeta))$}\,. \ee By (\ref{zeta-xi}) such a definition of $\hat A$
is $\zeta$-independent. $\hat A$ is skew-symmetric by (\ref{hatR})
and is skew-adjoint since \be \text{\rm Ran$(-\hat A\pm 1)$}
=L^2(\RE)\oplus L^2(\RE)\oplus \CO^n\oplus\CO^n \ee \n by
construction. \p Since we require $\hat A$ to be real, i.e. to
preserve the linear space (\ref{phys}), we have to restrict the
choice of $\Theta$ to real, skew-symmetric matrices. Off-diagonal
elements in the matrix $\Theta$ would correspond to non local couplings
between the pressure field and the oscillators. Since we are looking
for local interactions the only admissible choice for the
skew-symmetric matrix $\Theta$ is $\Theta=0$.

\n By (\ref{ran}) $(p,v,\uy,\uz)\in D(\hat A)$ if and only if
\begin{align}
p(x)&=p_\zeta(x)-\sum_{1\le i,j\le n}(\Gamma(\zeta)^{-1})_{ij}\,
(v_\zeta(s_j)-z_{\zeta\, j})\,\g'(x-s_i)\,,\\
v(x)&=v_\zeta(x)+\frac{\zeta}{a^2\rho_0}\,\sum_{1\le i,j\le n}(\Gamma(\zeta)^{-1})_{ij}\,
(v_\zeta(s_j)-z_{\zeta\,j})\,\g(x-s_i)\,,\\
\uy&=\uy_\zeta-S\sum_{1\le i,j\le n}(\Gamma(\zeta)^{-1})_{ij}\,
\,\frac{v_\zeta(s_j)-z_{\zeta\, j}}{K_i+\zeta^2M_i}\,\ue_i\,,\\
\uz&=\uz_\zeta-\zeta S\sum_{1\le i,j\le n}(\Gamma(\zeta)^{-1})_{ij}\,
\,\frac{v_\zeta(s_j)-z_{\zeta\, j}}{K_i+\zeta^2M_i}\,\ue_i\,,
\end{align}
\n with $(p_\zeta(x)\,,v_\zeta(x)\,,\uy_\zeta\,,\uz_\zeta)\in D(A)$.
\n Posing
\be
\hat A(p,v,\uy,\uz)\equiv(\hat A_1(p,v,\uy,\uz),\hat A_2(p,v,\uy,\uz),
\hat A_3(p,v,\uy,\uz),\hat A_4(p,v,\uy,\uz))\,,
\ee
The action of $\hat A$ on $(p,v,\uy,\uz)$ is given by
\begin{align}
[\hat A_1(p,v,\uy,\uz)](x)=&
-a^2\rho_0\frac{dv_\zeta}{dx}(x)-
\zeta\sum_{1\le i,j\le n}(\Gamma(\zeta)^{-1})_{ij}\,
(v_\zeta(s_j)-z_{\zeta\, j})\g'(x-s_i)\\
[\hat A_2(p,v,\uy,\uz)](x)=&-\frac{1}{\rho_0}\frac{dp_\zeta}{dx}(x)
+\frac{\zeta^2}{a^2\rho_0}\sum_{1\le i,j\le n}(\Gamma(\zeta)^{-1})_{ij}\,
(v_\zeta(s_j)-z_{\zeta\,j})\g(x-s_i)\\
\hat A_3(p,v,\uy,\uz)=&\uz_\zeta
-\zeta S\sum_{1\le i,j\le n}(\Gamma(\zeta)^{-1})_{ij}\,
\frac{v_\zeta(s_j)-z_{\zeta\, j}}{K_i+\zeta^2M_i}\,\ue_i\\
\hat A_4(p,v,\uy,\uz)=&-\sum_{1\le j\le
n}\frac{K_j}{M_j}\,y_{\zeta\,j}\,\ue_j
-\zeta^2 S\sum_{1\le i,j\le n}(\Gamma(\zeta)^{-1})_{ij}\,
\frac{v_\zeta(s_j)-z_{\zeta\, j}}{K_i+\zeta^2M_i}\,\ue_i\,.
\label{hatA4}
\end{align}
By the definitions of $D(\hat A)$ and $\Gamma(\zeta)$ one has
\be
\hat A_1(p,v,\uy,\uz)=-a^2\rho_0\,\frac{dv}{dx}
\ee
\be
\hat A_3(p,v,\uy,\uz)=\uz
\ee
and, defining
\be
\sigma_i:=p(s_i^+)-p(s_i^-)=\sum_{1\le j\le
n}(\Gamma(\zeta)^{-1})_{ij}\,
({v_\zeta(s_j)-z_{\zeta\, j}})\,,
\ee
formula (\ref{hatA4}) becomes
\begin{align}
\hat A_4(p,v,\uy,\uz)=&-\sum_{1\le j\le
n}\frac{K_j}{M_j}\,y_{\zeta\,j}\,\ue_j
-\zeta^2 S\sum_{1\le i\le n}\,
\frac{\sigma_i}{K_i+\zeta^2M_i}\,\ue_i=\\
=&-\sum_{1\le j\le
n}\left(\frac{K_j}{M_j}\,
y_{j}
+\frac{S}{M_j}\,\sigma_j\right)\,
\ue_j\,.
\end{align}
Then, posing
\be
p(x)=p_\zeta(x)-\sum_{1\le j\le n}\sigma_j\, \g'(x-s_j)=
p_0(x)+\frac{1}{2}\sum_{1\le j\le n}\sigma_j\,\sgn(x-s_j)\,,
\ee
one obtains
\begin{align}
[\hat A_2&(p,v,\uy,\uz)](x)=\nonumber\\
=&-\frac{1}{\rho_0}\frac{dp_0}{dx}(x)
-\sum_{1\le j\le
n}\frac{\sigma_j}{\rho_0}\left(\frac{d}{dx}\frac{|x-s_j|}
{2(x-s_j)}
-\left(-\frac{d^2}{dx^2}+\frac{\zeta^2}{a^2}\right)\g(x-y_j)\right)=\\
=&-\frac{1}{\rho_0}\,\frac{dp_0}{dx}\,.
\end{align}
Finally
\begin{align}
v(s_k)=&v_\zeta(s_k)+\frac{\zeta}{a^2\rho_0}\sum_{1\le i,j\le n}
(\Gamma(\zeta)^{-1})_{ij}\,
(v_\zeta(s_j)-z_{\zeta\,j})\,\frac{a}{\pm
2\zeta}\,e^{\mp\zeta|s_k-s_i|/a}=\\
=&v_\zeta(s_k)-\sum_{1\le i,j\le n}(\Gamma(\zeta)^{-1})_{ij}\,
(v_\zeta(s_j)-z_{\zeta\,j})\,\left((\Gamma(\zeta))_{ki}+
\frac{\zeta S\delta_{ki}}{K_i+\zeta^2M_i}\right)=\\
=&z_{\zeta\,k}-\zeta S
\sum_{1\le j\le n}(\Gamma(\zeta)^{-1})_{kj}\,
\frac{v_\zeta(s_k)-z_{\zeta\,k}}{K_k+\zeta^2M_k}=z_k\,.
\end{align}
\end{proof}
\n By the previous theorem the differential equation
\be
\frac{d}{dt}\,\Psi(t)=\hat A\Psi(t)
\label{problem}
\ee
is equivalent to the system of equations
\begin{align}
\frac{\partial p}{\partial t}&=-a^2\rho_0\,\frac{\partial v}{\partial x}\\
\frac{\partial v}{\partial t}&=-\frac{1}{\rho_0}\,\frac{\partial
p_0}{\partial x}
\equiv-\frac{1}{\rho_0}\,\left(\frac{\partial p}{\partial x}
-\sum_{1\le j\le n}\sigma_j\,\delta_{s_j}\right)\\
\frac{d\uy}{dt}&=\uz\\
\frac{d\uz}{dt}&=-\sum_{1\le j\le n}
\left(\frac{K_j}{M_j}\,y_j+\frac{S}{M_j}\,\sigma_j\right)\,\ue_j\,,
\end{align}
and the corresponding Cauchy problem generates the strongly continuous
unitary group of evolution exp$\,t\hat A$ on $L^2(\RE)\oplus
L^2(\RE)\oplus\CO^n\oplus\CO^n$ which preserves $D(\hat A)$. Here
$\delta_{s_j}$ denotes the Dirac mass at the point $s_j$ and
$\sigma_j$ (see (\ref{dom})) is the pressure jump at $s_j$.

\n It is worth noting that the only real, skew-adjoint extension of the free operator $A$ restricted to the space of the vectors $(p,v,\uy,\uz)$ such that $v(s_i,t)=z_i$ corresponds to the relevant physical coupling between the
pressure field and the oscillators.

\n The next result will be useful in the spectral analysis of $\hat
A$.

\begin{lemma}\label{lemmaGamma}
The matrix
\be
\Gamma_\pm(\lambda)^{-1}:=
\lim_{\ve\downarrow 0}\Gamma(\lambda\pm\ve)^{-1}
\ee
is well defined for any
$\lambda\in i\RE\backslash\{0\}$.
\end{lemma}
\begin{proof}
\n We give the proof only for the matrix $\Gamma_+(\lambda)$. The
proof for $\Gamma_-(\lambda)$ is analogous.

\n Let the matrix $\Gamma_+(\zeta)$ be the analytic continuation to
$\CO\backslash\cup_{j=1}^n\{\pm i\sqrt{K_j/M_j}\}$ of
$\Gamma(\zeta)$ defined for Re$\,\zeta> 0$ in (\ref{gamma}). Suppose
that $s_i>s_j$ if $i>j$, then \be
\Gamma_+(\zeta)=-\Pi(\zeta)-T(\zeta) \ee where $\Pi$ is the operator
\be \Pi=\left(\uphi^-(\zeta)\otimes\uphi^+(\bar\zeta)\right) \ee
with $\uphi^\pm(\zeta)=\sum_{i}\frac{e^{\pm\zeta
s_i/a}}{\sqrt{2a\rho_0}}\ue_i$. While $T(\zeta)$ is the upper
triangular matrix \be T(\zeta)_{ij}=
\begin{cases}
\frac{\zeta S\delta_{ij}}{K_i+\zeta^2M_i}+\frac{\sinh\zeta(s_i-s_j)}{a\rho_0}&i\leq j\\
0& i>j
\end{cases}
\ee

\n We use the formula
\begin{align}
\Gamma_+(\zeta)^{-1}=-\frac{1}{\Pi(\zeta)+T(\zeta)}&
=-\frac{1}{T(\zeta)}+\frac{1}{T(\zeta)}\,\Pi(\zeta)\,
\frac{1}{\Pi(\zeta)+T(\zeta)}=\\
&=-\sum_{n=0}^{\infty}\frac{(-1)^n}{T(\zeta)}\left(\Pi(\zeta)\,\frac{1}{T(\zeta)}\right)^n
\end{align}
valid for all $\zeta$ for which the series converges.

\n Matrix $T(\zeta)$ is invertible and its inverse $T(\zeta)^{-1}$ is a lower triangular matrix with $\left(T(\zeta)^{-1}\right)_{ii}=1/\left(T(\zeta)\right)_{ii}$. The eigenvalues of $T(\zeta)^{-1}$ are $1/\left(T(\zeta)\right)_{ii}$ and we can write
\be
T(\zeta)^{-1}=D(\zeta)\tilde T(\zeta)^{-1}D(\zeta)^{-1}
\ee
where $D(\zeta)$ is a unitary matrix, analytic for $\zeta\in\CO\backslash\{0\}$ and
\be
\left(\tilde
T(\zeta)^{-1}\right)_{ij}=\frac{1}{\left(T(\zeta)\right)_{ii}}\,
\delta_{ij}=\frac{K_i+\zeta^2M_i}{\zeta S}\,\delta_{ij}
\ee

\n We obtain for $\Gamma_+(\zeta)^{-1}$ the expression
\be
\Gamma_+(\zeta)^{-1}=-D(\zeta)\sum_{n=0}^{\infty}(-1)^n\left(\upsi(\zeta)\otimes\uchi(\zeta)\right)^n\tilde T(\zeta)^{-1}D(\zeta)^{-1}
\ee
with
\begin{align}
\left(\upsi(\zeta)\right)_i&=\frac{K_i+\zeta^2M_i}{\zeta S}\left(D(\zeta)^{-1}\uphi^-(\zeta)\right)_i\\
\left(\uchi(\zeta)\right)_i&=\left(D(\zeta)^{-1}\uphi^+(\bar\zeta)\right)_i
\end{align}
Then \be \Gamma_+(\zeta)^{-1}=
-\frac{1}{T(\zeta)}+\sum_{n=0}^\infty(-1)^n\left(\langle\uchi(\zeta),\upsi(\zeta)\rangle_{\CO^n}\right)^nD(\zeta)\upsi(\zeta)\otimes\uchi(\zeta)\tilde
T(\zeta)^{-1}D(\zeta)^{-1}\,. \ee For all $\zeta$ for which the
series converges one has \be
\Gamma_+(\zeta)^{-1}=-\frac{1}{T(\zeta)}+\frac{D(\zeta)\upsi(\zeta)\otimes\uchi(\zeta)\tilde
T(\zeta)^{-1}D(\zeta)^{-1}}{1+\langle\uchi(\zeta),\upsi(\zeta)\rangle_{\CO^n}}\,.
\label{sumseries} \ee

\n Consider the scalar product in $\CO^n$
\be
\left\langle\uchi(\zeta),\upsi(\zeta)\right\rangle_{\CO^n}
=\sum_{i=1}^n\overline{\left(D(\zeta)^{-1}\uphi^+(\bar\zeta)\right)}_i\,
\frac{K_i+\zeta^2M_i}{\zeta S}\,\left(D(\zeta)^{-1}\uphi^-(\zeta)\right)_i\,.
\label{chipsi}
\ee
Notice that, for $\lambda\in i\RE\backslash\{0\}$,
$\left\langle\uchi(\lambda),\upsi(\lambda)\right\rangle_{\CO^n}\in i\RE$ and
\begin{align}
-i \left\langle\uchi(\lambda),\upsi(\lambda)\right\rangle_{\CO^n}\to+\infty&\quad\textrm{for }\lambda\to+i\infty\\
-i \left\langle\uchi(\lambda),\upsi(\lambda)\right\rangle_{\CO^n}\to-\infty&\quad\textrm{for }\lambda\to i0^+\,.
\end{align}
Then there exists at least one point $\lambda\in i\RE$ in which
$\langle\uchi(\lambda),\upsi(\lambda)\rangle_{\CO^n}=0$. In a
neighborhood of this point the series converges and defines an analytic function. By (\ref{sumseries}) and (\ref{chipsi}) it is clear that  $\Gamma_+(\zeta)^{-1}$ exists for any
$\zeta\in\CO\backslash\{0\}$. The same  relations
show that one can put $\Gamma_+(\zeta)^{-1}:=0$ if
$\zeta=i\sqrt{K_j/M_j}$, $j=1,\dots,n$.
\end{proof}

\n The following theorem completely characterizes the spectrum of $\hat A$.
\begin{theorem}The essential spectrum of $\hat A$ is purely absolutely continuous and
\be
\sigma_{ess}(\hat A)=\sigma_{ac}(\hat A)=i\RE\,,\qquad\sigma_{pp}(\hat A)=\left\{0\right\}\,.
\ee
\n  Any vector of the kind
\be
\left(\frac{1}{2}\sum_{1\le j\le n}\sigma_j\,\sgn(x-s_j),\,0,\,-\sum_{1\le
j\le n}\frac{S}{K_j}\,\sigma_j\,\ue_j,\, \underline 0\right)\,,
\label{eigenveczero}
\ee
with
\be
\sum_{1\le j\le n}\sigma_j=0\,,
\label{sigmacond}
\ee
is an eigenvector corresponding to the $(n-1)$-fold degenerate eigenvalue $\lambda=0.$

\n The generalized eigenfunctions  $\hat\Phi^\pm(\lambda)$ corresponding
to the point of the absolutely continuous spectrum relative to right $(+)$ and left $(-)$ incidence are given by
\be
\hat\Phi^\pm(\lambda,x)=\left(\hat\phi_p^\pm(\lambda,x),\hat\phi_v^\pm(\lambda,x),
\hat\phi_{\uy}^\pm(\lambda),
\hat\phi_{\uz}^\pm(\lambda)\right)\qquad \lambda\in i\RE
\ee
\begin{align}
\hat\phi_p^\pm(\lambda,x)&=Ce^{\pm\lambda x/a}
\mp\frac{C}{2a\rho_0}\sum_{1\le i,j\le n}
(\Gamma_+(\lambda)^{-1})_{ij}
e^{\pm\lambda s_j/a}\sgn(x-s_i)e^{-\lambda|x-s_i|/a}\\
\hat\phi_v^\pm(\lambda,x)&=\mp C\frac{e^{\pm\lambda x/a}}{a\rho_0}
\mp\frac{C}{2a^2\rho_0^2}\sum_{1\le i,j\le n}
(\Gamma_+(\lambda)^{-1})_{ij}
e^{\pm\lambda s_j/a}e^{-\lambda|x-s_i|/a}\\
\hat\phi_{\uy}^\pm(\lambda)&=\pm\frac{SC}{a\rho_0}
\sum_{1\le i,j\le n}
(\Gamma_+(\lambda)^{-1})_{ij}
\frac{e^{\pm\lambda s_j/a}}{K_i+\lambda^2M_i}\ue_i\\
\hat\phi_{\uz}^\pm(\lambda)&=\pm\frac{\lambda SC}{a\rho_0}\sum_{1\le i,j\le
n}
(\Gamma_+(\lambda)^{-1})_{ij}
\frac{e^{\pm\lambda s_j/a}}{K_i+\lambda^2M_i}\ue_i
\end{align}
with $C=\sqrt{a\rho_0/(4\pi)}$.
\end{theorem}

\begin{proof}
\n For $\zeta \in \rho(A)\cap\rho(\hat A)$, $(-\hat
A+\zeta)^{-1}-(-A+\zeta)^{-1}$ is of finite rank, then from Weyl's
criterion (see e.g. \cite{RS4} Theorem XIII.14) one has
$\sigma_{ess}(\hat A)=\sigma_{ess}(A)=i\RE$.
Moreover, by Birman-Kato invariance principle, the wave operators
$\Omega_\pm(\hat A,A)$ exist and are complete (see e.g. \cite{RS3},
Corollary 2 to Theorem XI.11). Thus $\sigma_{ac}(\hat A)=\sigma_{ac}(A)$.

\n Let $\hat\mu_\Psi^{sc}$ be the singular continuous part of the
spectral measure on $i\RE$ corresponding to $\hat A$ and $\Psi$. Since
$\|\breve G(\zeta)\Psi\|<\infty$ for all
$\zeta\in\CO\backslash\sigma_{pp}(A)$ and for all $\Psi\in D$,
\be
D:=\left\{\Psi\equiv(p,v,\uy,\uz)\,:\, p\in L^1(\RE)\cap
L^2(\RE)\,,\quad v\in L^1(\RE)\cap L^2(\RE) \right\}\,,
\ee
by Lemma \ref{lemmaGamma} and \cite{RS4}, Theorem XIII.19, one has
supp$\,\hat\mu_\Psi^{sc}\subseteq \left\{0\right\}\cup\sigma_{pp}(A)$
i.e. supp$\,\hat\mu_\Psi^{sc}=\emptyset$ since $\hat\mu_\Psi^{sc}$ has
no atoms by its definition. Since $D$ is dense this gives
$\sigma_{sc}(\hat A)=\emptyset$.

\n One can check that any vector $\Psi$ of the kind (\ref{eigenveczero}) is in the domain of $\hat A$ and solves the equation $\hat A\Psi=0$. The degeneration of eigenvalue $\{0\}$ follows from condition  (\ref{sigmacond}).

\n Suppose now $\lambda\in i\RE\backslash\{0\}$ and consider the equation $\hat A\Psi=\lambda\Psi$. This produces, if $\Psi\equiv(p,v,\uy,\uz)$, the equation
\be
v''-\frac{\lambda^2}{a^2}\,v=-\frac{\lambda}{a^2\rho_0}\sum_{1\le j\le n}\sigma_j\delta_{s_j}\,,
\ee
with $\sigma_i\in\CO$, $i=1,\dots, n$, which has no square integrable solution.

\n The expression for the generalized eigenfunctions is
a consequence of the Stone's formula (see e.g. \cite{RS1}, Theorem
VII.13)
which gives the generalized expansion formula
\be
\Psi=\textrm{s -}\lim_{a\downarrow-\infty,\,b\uparrow\infty}\textrm{s
-}
\lim_{\ve\downarrow 0}\,\frac{1}{2\pi}\int_a^b
[\hat R(\lambda+\ve)-\hat R(\lambda-\ve)]\Psi\, d\lambda\,.
\ee
\end{proof}

\n In the following lemma the asymptotic behavior of the oscillations
of the thin walls is characterized. It is proved that the oscillators relax
(as $|t|\to\infty$) towards their equilibrium positions
for any initial data orthogonal to
the eigenspace relative to eigenvalue zero. For example this is true
for any initial datum of the kind $\Psi_0\equiv(p,v,\underline 0,
\uz)$ where the support
of $p$ is outside the interval containing the points $s_1,\dots, s_n$
which denote the equilibrium position of the walls.
\begin{lemma} Given $\Psi_0$ orthogonal to the eigenspace relative to eigenvalue zero,
let us denote by $\left(\uy(t),\uz(t)\right)$ the projection onto
$\CO^{n}\oplus \CO^n$ of
$e^{t\hat A}\Psi_0$. Then
$$
\lim_{|t|\to\infty}\,\|\uy(t)\|_{\CO^n}=0\quad\text{and}\quad
\lim_{|t|\to\infty}\,\|\uz(t)\|_{\CO^n}=0\,.
$$
\end{lemma}
\begin{proof} Let $\hat P(dk)$ be the projection-valued measure
corresponding to the self-adjoint
operator $-i\hat A$. Since $\Psi_0$ is in the absolutely continuous
subspace, for any $\Psi$ the bounded complex measure
$\langle\langle\Psi,\hat P(dk)\Psi_0\rangle\rangle$
is absolutely continuous with respect to
Lebesgue measure and hence its density belongs to $L^1(\RE)$. Thus, by
the spectral theorem and Riemann-Lebesgue lemma,
\be
\lim_{|t|\to\infty}\,\langle\langle\Psi,e^{t\hat
A}\Psi_0\rangle\rangle=
\lim_{|t|\to\infty}\,\int_\RE e^{-itk}\,
\langle\langle\Psi,\hat P(dk)\Psi_0\rangle\rangle
=0\,.
\ee
By taking $\Psi=(0,0,\ue_i,\underline 0)$ and
$\Psi=(0,0,\underline 0,\ue_i)$, $i=1,\dots,n$, one then obtains
\be
\lim_{|t|\to\infty}\,y_i(t)=0\quad\text{and}\quad
\lim_{|t|\to\infty}\,z_i(t)=0\,.
\ee
\end{proof}

\n In order to obtain more precise estimate on the asymptotic behavior of solutions of equation (\ref{problem}), for particular initial conditions, a detailed analysis of $\Gamma(\lambda)^{-1}$ is required. For example in specific cases one can prove existence of frequencies which are totally transmitted by the array of oscillators.

\section{\label{secthree}Kronig-Penney model in acoustics}

\n It is possible to extend the previous construction to the case of
an array of  infinitely many oscillators. We prove that in the case
of a periodic array of identical oscillators the energy spectrum
shows a band structure.

\n As a first step we define the operator $\hat A$ introduced in section \ref{ssingular} when
$\S=\left\{s_1,s_2,\dots\right\}$ is a  denumerable
set such that
\be
d:=\inf_{i\not= j}|s_i-s_j|>0\quad i,j\in \N.
\ee
Defining the linear map
\be
\tau(p,v,\uy,\uz):=\sum_{j=1}^\infty(v(s_j)-z_j)\,\ue_j\,,
\ee
where $\left\{\ue_j\right\}_1^\infty$ is the usual complete
orthonormal system for $\l^2$, one has the following
\begin{lemma} $\tau$ is bounded as a map on
$H^1(\RE)\oplus H^1(\RE)\oplus\l^2\oplus\l^2$ to $\l^2$.
\end{lemma}
\begin{proof} We will follow closely \cite{AGH-KH}. Let $\left\{I_j\right\}_1^\infty$ be a partition of
$\RE$ and let $K(x-y)$ be the kernel of $(-\Delta+1)^{-1/2}$.
Since
\be
v(x)=\sum_{j=1}^\infty\int_{I_j}K(x-y)\,[(-\Delta+1)^{1/2}v](y)\,dy\,,
\ee
to prove the lemma amounts to show that the infinite matrix
\be
M_{ij}:=\left(\int_{I_j}K(x-s_i)^2\,dx\right)^{1/2}
\ee
corresponds to a bounded linear operator $M$ on $\ell^2$.
By Lemma C.3 in \cite{AGH-KH}, one has
\begin{align}
\|M\|^2_{\l^2,\l^2}\le&\sup_{i}
\sum_{j=1}^\infty\left(\int_{I_j}K(x-s_i)^2\,dx\right)^{1/2}
 \sup_{j}\sum_{i=1}^\infty\left(\int_{I_j}K(x-s_i)^2\,dx\right)^{1/2}
\,.
\end{align}
Since
$$
\frac{1}{\sqrt
x}=\frac{2}{\pi}\int_0^\infty\frac{d\mu}{x+\mu^2}\,,\qquad x>0\,,
$$
by functional calculus one has
\be
K(x-y)
=\frac{1}{\pi}\int_0^\infty
\frac{e^{-\sqrt{1+\mu^2}\,|x-y|}}{\sqrt{1+\mu^2}}\,d\mu \,.
\ee
By taking $I_j=[s_j-\epsilon_j,s_j+\delta_j)$, where $\epsilon_j$ is
one half the distance between $s_j$ and the preceding point and
$\delta_j$ is one half the distance between $s_j$ and the successive
point, a straightforward calculation leads to
\begin{align}
&\int_{I_j}K(x-s_i)^2\,dx\leq
\frac{2}{d^2\pi^2}\,{e^{-|s_i-s_j|/\sqrt 2}}\,,
\end{align}
from where the estimate $\|M\|_{\l^2,\l^2}<+\infty$ follows immediately.
\end{proof}
\n The construction proceeds now along
the same lines as in the case of a finite set of points. We state the final
result:
\begin{theorem} Let $\{K_j\}_1^\infty$ $\{M_j\}_1^\infty$, $K_j>0$,
$M_j>0$ be in $\l^\infty$ and suppose that $\{K_j/M_j\}_1^\infty$ and
$\{1/M_j\}_1^\infty$ are in $\l^\infty$ too. The linear operator
\be
\hat A:D(\hat A)\subset L^2(\RE)\oplus
L^2(\RE)\oplus\l^2\oplus\l^2\to L^2(\RE)\oplus L^2(\RE)\oplus\l^2\oplus\l^2\,,
\ee
\be
\begin{aligned}
D(\hat A)
=\{&(p,v,\uy,\uz)\,:\, p\in L^2(\RE)\cap H^1(\RE\backslash \S),\, v\in
H^1(\RE),\  \uy\in\l^2,\, \uz\in\l^2,\\ &p(s_i^+)-p(s_i^-)=\sigma_i,\
v(s_j)=z_j,\ \usigma\in \l^2\}\,,
\end{aligned}
\ee
\be
\begin{aligned}
&\hat A(p,v,\uy,\uz):=
\\&:=\left(-a^2\rho_0\,\frac{dv}{dx},\,
-\frac{1}{\rho_0}\,\frac{dp_0}{dx},\,\uz,\,
-\sum_{j=1}^\infty\left(\frac{K_j}{M_j}\,y_j+\frac{S}{M_j}\,\sigma_j\right)\,\ue_j\right)
\end{aligned}
\ee
is real and skew-adjoint. Here $p_0\in \bar H^1(\RE)$,
\be
p_0(x):=p(x)-\frac{1}{2}\sum_{j=1}^\infty\sigma_j\,
\sgn(x-s_j)\,,
\ee
denotes the regular part of $p$. The resolvent of $\hat A$ is given by
\be
(-\hat A+\zeta)^{-1}=
(-A+\zeta)^{-1}+\sum_{i,j=1}^\infty\left(\Gamma(\zeta)^{-1}\right)_{ij}\,
G^i_\zeta\otimes
\breve G^j_{\bar\zeta}\,,\qquad \zeta\in\CO\backslash i\RE\,.
\ee
\label{theoinfty}
\end{theorem}

\n Now we can proceed to the study of a periodic system. We use the same
notation of \cite{AGH-KH}.

\n In this case $\S$ will be the ``Bravais'' lattice,
\be
\S=\{nL:n\in\Z\}\,, \quad L>0\,,
\ee
\n and $\hat\S$ the ``Brillouin'' zone,
\be
\hat\S=\left[-\frac{b}{2},\frac{b}{2}\right)\,,\quad b=\frac{2\pi}{L}\,.
\ee

\n We consider a Hilbert  space $\HH$  on $L^2\oplus L^2\oplus\l^2\oplus\l^2$
 in which the scalar product is defined by
\be
\frac{1}{a^2\rho_0}\langle p_1,p_2\rangle+
\rho_0\langle v_1,v_2\rangle+
\frac{K}{S}\langle\uy_1,\uy_2\rangle+
\frac{M}{S}\langle\uz_1,\uz_2\rangle
\ee
\n where $\langle\cdot,\cdot\rangle$ represents either the usual scalar
product in $L^2$, when concerning pressure and velocity fields, or the
usual scalar product in $\l^2$, for $\uy$ and $\uz$.

\n $M$, $K$ and $S$ are positive constants representing the mass of oscillating walls, the elastic constant of the
springs  and the area of the transverse section of
the pipe.

\n The Hilbert space $\HH$ can be decomposed as
\begin{align}
\HH&=\W^{-1}\tHH(\hat\S,b^{-1}d\te;L^2([-L/2,L/2))\oplus
L^2([-L/2,L/2))\oplus\CO\oplus\CO)\\
&=\W^{-1}\int_{[-b/2,b/2)}^{\oplus}\frac{d\te}{b}\Bigg(
L^2([-L/2,L/2))\oplus
L^2([-L/2,L/2))\oplus\CO\oplus\CO\Bigg)
\end{align}
\n where
\be
\W:\,\HH\to\tHH(\hat\S ,b^{-1}d\te;L^2([-L/2,L/2))\oplus
L^2([-L/2,L/2))\oplus\CO\oplus\CO)
\ee
\be
\W(p,v,\uy,\uz)\equiv\left((\WW p)(\te,\nu),(\WW
v)(\te,\nu),(\WW \uy)(\te),(\WW \uz)(\te)\right)
\ee
\begin{align}
&(\WW p)(\te,\nu)\equiv\tilde{p}(\te,\nu)=\sum_{n\in\Z}e^{in\te
L}p(\nu+nL)\\
&(\WW v)(\te,\nu)\equiv\tilde{v}(\te,\nu)=\sum_{n\in\Z}e^{in\te
L}v(\nu+nL)\\
&(\WW\uy)(\te)\equiv\tilde{y}(\te)=\sum_{n\in\Z}e^{in\te
L}y_n\\
&(\WW\uz)(\te)\equiv\tilde{z}(\te)=\sum_{n\in\Z}e^{in\te
L}z_n\qquad\nu\in[-L/2,L/2)\;,\quad\te\in[-b/2,b/2)
\end{align}
\n and
\be
\W^{-1}:\,\tHH(\hat\S,b^{-1}d\te;L^2([-L/2,L/2))\oplus
L^2([-L/2,L/2))\oplus\CO\oplus\CO)\to\HH
\ee
\be
\begin{aligned}
\W^{-1}&(\tilde{p},\tilde{v},\tilde\uy,\tilde\uz)\equiv\\
&\equiv\left((\WW^{-1}\tilde{p})(\nu+nL),(\WW^{-1}\tilde{v})(\nu+nL)
,\{(\WW^{-1}\tilde{y})_n\},\{(\WW^{-1}\tilde{z})_n\}\right)
\end{aligned}
\ee
\begin{align}
&(\WW^{-1}\tilde{p})(\nu+nL)=b^{-1}\int_{-b/2}^{b/2}d\te e^{-in\te L}
\tilde{p}(\te,\nu)\\
&(\WW^{-1}\tilde{v})(\nu+nL)=b^{-1}\int_{-b/2}^{b/2}d\te e^{-in\te L}
\tilde{v}(\te,\nu)\\
&(\WW^{-1}\tilde{y})_n=b^{-1}\int_{-b/2}^{b/2}d\te e^{-in\te
L}\tilde{y}(\te)\\
&(\WW^{-1}\tilde{z})_n=b^{-1}\int_{-b/2}^{b/2}d\te e^{-in\te L}\tilde{z}(\te)
\qquad\nu\in[-L/2,L/2)\;,\quad n\in\Z\,.
\end{align}
The scalar product in $L^2([-L/2,L/2))\oplus
L^2([-L/2,L/2))\oplus\CO\oplus\CO$ is defined by
\be
\frac{1}{a^2\rho_0}\langle \tp_1,\tp_2\rangle_{L/2}+
\rho_0\langle \tv_1,\tv_2\rangle_{L/2}+
\frac{K}{S} \bar{\ty}_1\ty_2+
\frac{M}{S} \bar{\tz}_1\tz_2
\label{scprtheta}
\ee
\n where $\langle\cdot,\cdot\rangle_{L/2}$ indicates the usual scalar
product in $L^2([-L/2,L/2))$.

\n From Theorem \ref{theoinfty} we obtain the following
\begin{corollary}
The linear operator
\be
\hat A:D(\hat A)\subset \HH\to\HH
\ee
\be
\begin{aligned}
D(\hat A)=\Big\{&(p,v,\uy,\uz)\,:p\in L^2(\RE)\cap
H^1(\RE\backslash\S)\,,\;v\in H^1(\RE)\,,\uy\in\l^2\,,\;\uz\in\l^2\,,\\
&p(nL^+)-p(nL^-)=\sigma_n\,, v(nL)=z_n\;\forall n\in\Z\,,\;\usigma\in\l^2\Big\}
\end{aligned}
\ee
\be
\hat
A(p,v,\uy,\uz):=\left(-a^2\rho_0\frac{dv}{dx},-\frac{1}{\rho_0}\frac{dp_0}{dx},
\uz,-\frac{K}{M}\uy-\frac{S}{M}\usigma\right)\,,
\ee

\n where the regular part of $p(x)$, denoted with $p_0\in\bar H^1(\RE)$, is
\be
p_0(x)=p(x)-\frac{1}{2}\sum_{n\in\Z}\sigma_n\,\sgn(x-nL)\,,
\ee
\n is real and skew-adjoint.
\end{corollary}

\n We want to study  the spectral structure of $\hat A$. To this aim we introduce the family  of operators
$\hat A(\theta)$
\be
\begin{aligned}
\hat A(\te):D(\hat A(\te))&\subset L^2((-L/2,L/2))\oplus
L^2((-L/2,L/2))
\oplus\CO\oplus\CO\\
&\to L^2((-L/2,L/2))\oplus
L^2((-L/2,L/2))
\oplus\CO\oplus\CO
\end{aligned}
\ee
\be
\begin{aligned}
D(\hat A(\te))=\Bigg\{&(\tp(\te),\tv(\te),\ty(\te),\tz(\te))\,:\\
&\tp(\te)\in H^1((-L/2,L/2)\backslash\{0\})\,,\;
\tv(\te)\in H^1((-L/2,L/2))\,,\\
&\ty(\te)\in\CO\,,\;\tz(\te)\in\CO\\
&\tp(\te,0^+)-\tp(\te,0^-)=\tsig(\te)\,,\\
&\tv(\te,0)=\tz(\te)\,,\;\tsig(\te)\in\CO\,,\\
&\tp\left(\te,-\frac{L}{2}^+\right)
=e^{i\te L}\tp\left(\te,\frac{L}{2}^-\right),\\
&\tv\left(\te,-\frac{L}{2}^+\right)=
e^{i\te L}\tv\left(\te,\frac{L}{2}^-\right)\Bigg\}\,;\quad
\forall\te\in\left[-\frac{b}{2},\frac{b}{2}\right)
\end{aligned}
\ee
\be
\begin{aligned}
\hat
A(\te)(\tp(\te)&,\tv(\te),\ty(\te),\tz(\te)):=
\\&:=\left(-a^2\rho_0\frac{d\tv(\te)}{d\nu},-\frac{1}{\rho_0}\frac{d\tp_0(\te)}{d\nu},
\tz(\te),-\frac{K}{M}\ty(\te)-\frac{S}{M}\tsig(\te)\right)
\end{aligned}
\ee
\n where $\tp_0(\te)\in H^1(\RE)$ is the regular part of $\tp(\te)$
\be
\tp_0(\te,\nu)=\tp(\te,\nu)-\frac{1}{2}\tsig(\te)\,\sgn(\nu)\,.
\ee

\n Boundary conditions for $\tp(\te,\nu)$ and $\tv(\te,\nu)$ in $\nu=0$ and $\nu=\pm L/2$ are such that all
operators in this family are  skew-adjoint with respect to the scalar
product (\ref{scprtheta}).

\n The operator $\hat A$ is related to $\hat A(\te)$ by the relation (see \cite{AGH-KH})
\be
\W\hat A\W^{-1}=\int_{[-b/2,b/2)}^{\oplus}\frac{d\te}{b}\hat A(\te)\,.
\ee

\begin{figure}[h!]
     \centering
     \subfigure[With non degenerate eigenvalues. $M/M_g=0.5$, $\omega_o/\omega_g=1.2$]{
          \label{nondeg}
          \includegraphics[width=.50\textwidth]{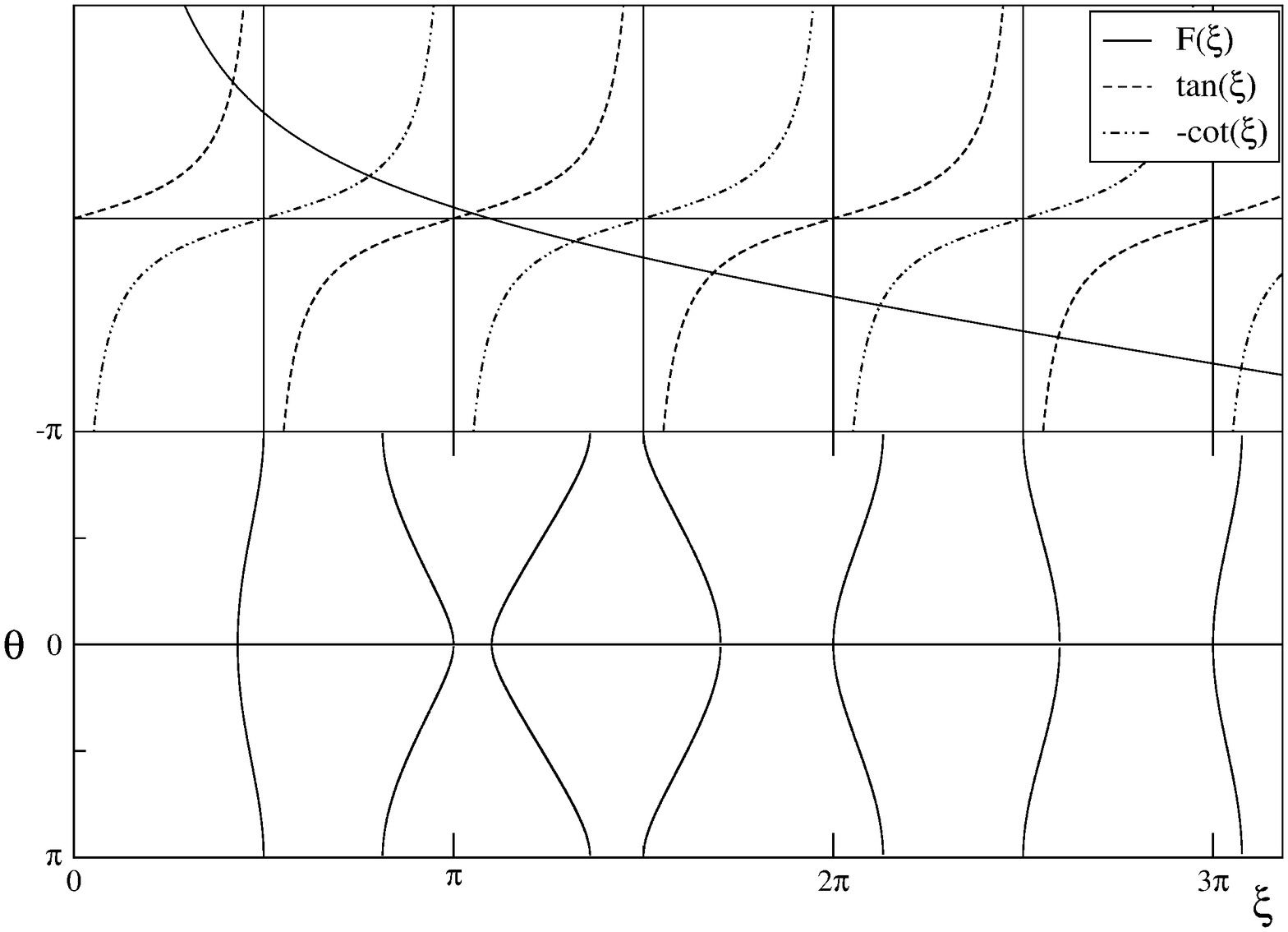}}
\hspace{.05in}
     \subfigure[With one degenerate eigenvalue. $M/M_g=0.5$, $\omega_o/\omega_g=1$]{
          \label{deg}
          \includegraphics[width=.50\textwidth]{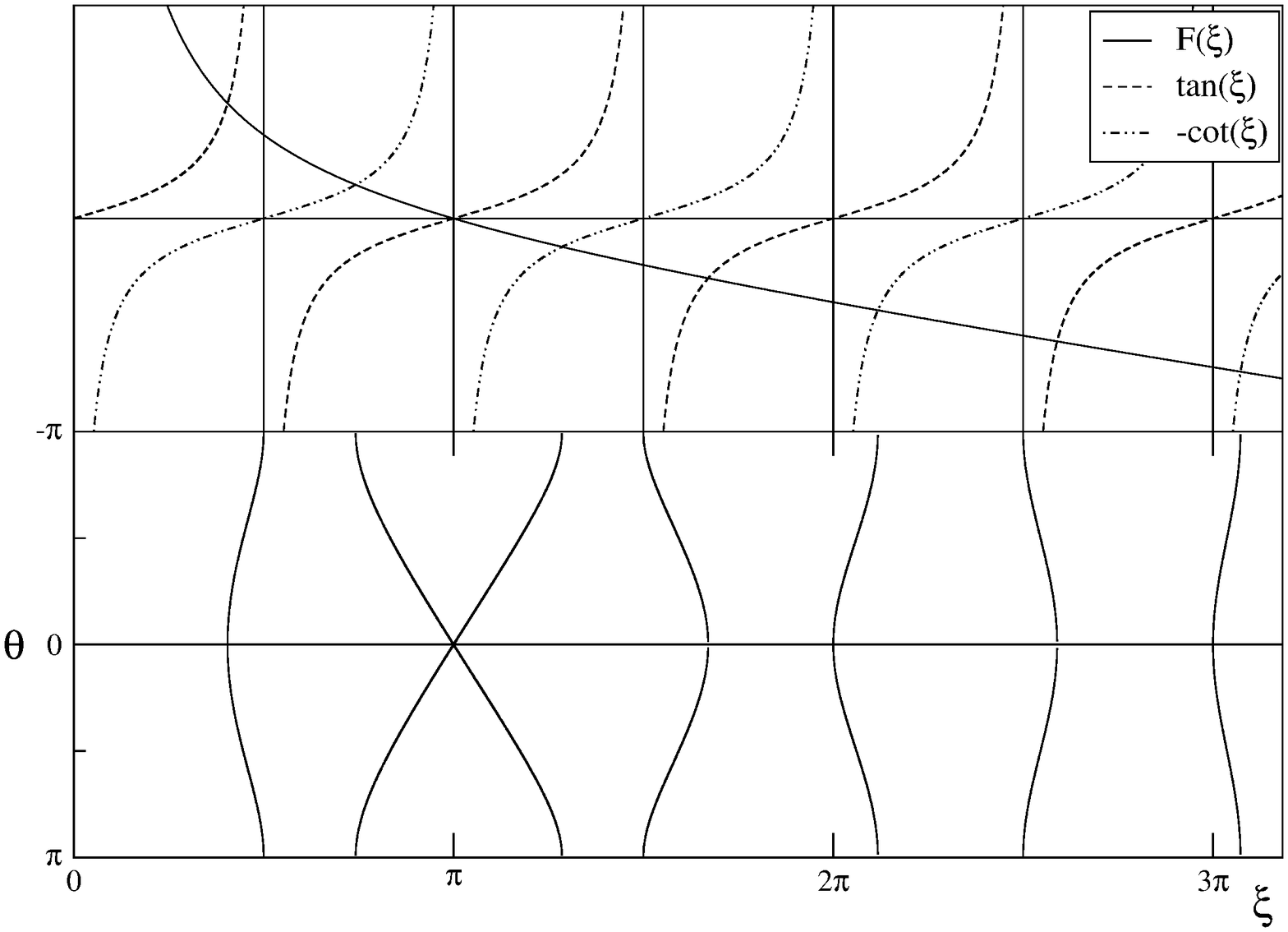}}
\caption{\label{bande}The upper part of figures shows the graphical
solution of equations (\ref{tetazero}) and (\ref{tetab2}). The lower
part shows the band structure.}
\end{figure}

\n The spectrum of $\hat A(\te)$ is described by the following
\begin{theorem}
Let $\te\in[-b/2,b/2)$ then the spectrum of $\hat A(\te)$ is  purely discrete, in particular
its eigenvalues $E_n(\te)$  are given by
\be
E_n(\te)=\lambda_n(\te)=2i\xi_n(\te)\frac{a}{L}\;;\quad n\in\Z\,,\;\xi_n(\te)\in\RE
\label{En}
\ee
\n where $\xi_n(\te)$ are the real solutions of
\be
\sin\xi\left[\sin\xi-F(\xi)\cos\xi\right]\cos^2\frac{\te L}{2}
=\cos\xi\left[\cos\xi+F(\xi)\sin\xi\right]\sin^2\frac{\te L}{2}
\label{evs}
\ee
\be
F(\xi)=\frac{M}{M_g}\left(\pi^2\frac{\omega_o^2}{\omega_g^2}\frac{1}{\xi}-
\xi\right)\;;\quad M_g=\rho_0 SL,\;\omega_o^2=\frac{K}{M},\;
\omega_g=2\pi\frac{a}{L}.
\ee

\n The corresponding eigenfunctions are
\be
\Phi_n(\te,x)=(\tp_n(\te,\nu),\tv_n(\te,\nu),\ty_n(\te),\tz_n(\te))\;;\quad
n\in\Z,\;
\te\in[-b/2,b/2)
\ee
\begin{align}
\tp_n(\te,\nu)=&C_n\Bigg[\left(\sin\left(\xi_n-\frac{\te L}{2}\right)-
                    F(\xi_n)\cos\left(\xi_n-\frac{\te L}{2}\right)\right)
                    \cos\frac{2\xi_n}{L}\nu+\nonumber\\
            &-i\sin\left(\xi_n-\frac{\te L}{2}\right)
                    \left(\sin\frac{2\xi_n}{L}\nu-
                    F(\xi_n)\frac{|\nu|}{\nu}\cos\frac{2\xi_n}{L}\nu\right)\Bigg]\\
\tv_n(\te,\nu)=&-\frac{iC_n}{a\rho_0}
                   \Bigg[\left(\sin\left(\xi_n-\frac{\te L}{2}\right)-
                    F(\xi_n)\cos\left(\xi_n-\frac{\te L}{2}\right)\right)
                    \sin\frac{2\xi_n}{L}\nu+\nonumber\\
            &+i\sin\left(\xi_n-\frac{\te L}{2}\right)
                    \left(\cos\frac{2\xi_n}{L}\nu+F(\xi_n)\sin\frac{2\xi_n}{L}|\nu|
                  \right)\Bigg]\\
\ty_n(\te)=&-i\frac{C_nL}{a^2\rho_0\xi_n}\sin\left(\xi_n-\frac{\te L}{2}\right)\\
\tz_n(\te)=&\frac{C_n}{a\rho_0}\sin\left(\xi_n-\frac{\te L}{2}\right)
\end{align}

\n For $\te\in[-b/2,b/2)$ zero is an eigenvalue with eigenfunction
\be
\Psi_0=\left(C_0\left(\cos\frac{\te L}{2}-i\sin\frac{\te L}{2}\sgn(\nu) \right),0,2iC_0\frac{S}{K}\sin\frac{\te L}{2},0\right)
\label{efzero}
\ee

\n Moreover the following chain of inequalities holds
\be
\begin{aligned}
&0< E_1(0)<E_1(-b/2)\leq E_2(-b/2)<E_2(0)\leq E_3(0)<E_3(-b/2)\leq\\
&\leq E_4(-b/2)<E_4(0)\leq E_5(0)<E_4(-b/2)\leq E_5(-b/2)<\dots
\label{chain}
\end{aligned}
\ee

\n In general the eigenvalues $E_n(\te)$ are all  distinct and  non
degenerate. If $\omega_o/\omega_g=n/2$ with $n\in\N$ there is just
one two fold degenerate eigenvalue equal to $n\pi/2$, such
eigenvalue corresponds to $\te=0$ for $n$ even and to $|\te|=b/2$
for $n$ odd.

\n If $E(\te)$ is an eigenvalue then $-E (\te)$
is an eigenvalue.

\n Given $\te\in[-b/2,b/2)$ the following relation holds
\be
E_n(-\te)=E_n(\te)\,.
\label{relation}
\ee
\end{theorem}
\begin{proof}
\n Eigenvalues and eigenfunctions (\ref{En})-(\ref{efzero}) are
given by direct computation. We solve the system of equations \be
\hat
A(\te)(\tp(\te),\tv(\te),\ty(\te),\tz(\te))=\lambda(\tp(\te),\tv(\te),\ty(\te),\tz(\te))\qquad\lambda\in
i\RE \ee

\n with the condition $\tv(\te,0)=\tz(\te)$, the solution reads
\begin{align}
\tp(\te,\nu)=&C(\xi)\cos\frac{2\xi\nu}{L}+D(\xi)\left[\sin\frac{2\xi\nu}{L}-F(\xi)\sgn(\nu)\cos\frac{2\xi\nu}{L}\right]\\
\tv(\te,\nu)=&\frac{C(\xi)}{ia\rho_0}\sin\frac{2\xi\nu}{L}-\frac{D(\xi)}{ia\rho_0}\left[\cos\frac{2\xi\nu}{L}+F(\xi)\sin\frac{2\xi|\nu|}{L}\right]
\end{align}

\n where $\xi=-iL\lambda/(2a)\in\RE$, $C(\xi)$ and $D(\xi)$ are two unknown functions of $\xi$. To determine $C(\xi)$ and $D(\xi)$ we have to take into account the boundary conditions
\be
\left\{
\begin{aligned}
\tp\left(\te,-\frac{L}{2}^+\right)=&e^{i\te L}\tp\left(\te,\frac{L}{2}^-\right)\\
\tv\left(\te,-\frac{L}{2}^+\right)=&e^{i\te L}\tv\left(\te,\frac{L}{2}^-\right)
\label{system}
\end{aligned}
\right.
\ee

\n This system has only the trivial solution $C(\xi)=0$ and
$D(\xi)=0$ for the values of $\xi$ for which the determinant  of the
matrix of the coefficients of the system is zero. The condition that
the determinant is zero implies equation (\ref{evs}) for the
eigenvalues. For $\xi$ satisfying condition (\ref{evs}) the
solutions of the system of dependent equations (\ref{system}) give
the eigenfunctions.

\n For $\te=0$ and $\te=-b/2$ relation (\ref{evs}) becomes
\begin{align}
\tan\xi&=0&&\textrm{or}&
\tan\xi&=F(\xi)=
\frac{M}{M_g}\left(\pi^2\frac{\omega_o^2}{\omega_g^2}\frac{1}{\xi}
-\xi\right)\;;\quad\te=0\label{tetazero}\\
\cot\xi&=0&&\textrm{or}&
-\cot\xi&=F(\xi)=
\frac{M}{M_g}\left(\pi^2\frac{\omega_o^2}{\omega_g^2}\frac{1}{\xi}
-\xi\right)\;;\quad\te=-b/2\label{tetab2}
\end{align}

\n Graphic solutions of the transcendental equations (\ref{tetazero}) and (\ref{tetab2}) are given in the upper part of figures
\ref{nondeg} and \ref{deg}. The chain of inequalities (\ref{chain}) follows by the
monotone behavior of  $F(\xi)$.

\n Degeneration of eigenvalues for $\omega_o/\omega_g=n/2$, the fact that $-E(\te)$ is an eigenvalue if $E(\te)$ is an eigenvalue and relation (\ref{relation}) follow directly by equation (\ref{evs}) and by $F(\xi)=-F(-\xi)$.
\end{proof}


\vspace{0.1in}

\n One can show that there is a band structure writing equation
  (\ref{evs}) as
\be
\tan^2\frac{\te
L}{2}=\tan{\xi}\left[\frac{\tan\xi-F(\xi)}{1+F(\xi)\tan\xi}
\right]
\label{eqbande}
\ee

\n It is possible to find solutions of equation (\ref{eqbande}) only for values
of $\xi$ such that the r.h.s. is positive. In the lower part of
 figures \ref{deg} and \ref{nondeg}  the resulting band structure is shown. The figures
clearly show that the width of the gaps is connected to the structure
of the spectrum. In particular  figure \ref{deg} shows that when
there is a degenerate eigenvalue, $\omega_o/\omega_g=n\pi/2$ with
$n\in\N$, a gap disappears because of the overlapping of two bands.

\n The bandwidth increases, when the ratio $M/M_g$ decreases.


\newpage

\begin{thebibliography}{99}

\bibitem[AGH-KH]{AGH-KH} S. Albeverio, F. Gesztesy, R. H\o egh-Krohn, H. Holden,
\emph{Solvable Models in Quantum Mechanics: second edition}, AMS
Chelsea Publ. (2005).

\bibitem[D]{D} P.A.M. Dirac, \emph{Classical Theory of Radiating Electrons}, Proc. R. Soc. London, Ser.
A, {\bf 167}, 148-169 (1938).

\bibitem[GG]{GG} V.I. Gorbachuk, M.L. Gorbachuk, \emph{Boundary Value
Problems for Operator Differential Equations}, Kluwer
Acad. Publ. (1991).

\bibitem[GS]{GS} D.J. Griffiths, C.A. Steinke, \emph{Waves in Locally Periodic Media}, Am. J. Phys., {\bf 69}, No. 2, 137-154 (2001).

\bibitem[IW]{IW} L. Infeld, P.R. Wallace, \emph{The Equation of Motion
in Electrodynamics}, Phys. Rev., {\bf 57}, 797-806 (1940).

\bibitem[K]{K} J. Kijowski, \emph{Electrodynamics of Moving
Particles},
Gen. Rel. Grav., {\bf 26}, 167-201 (1994).

\bibitem[M]{M} M. Marino, \emph{Classical Electrodynamics of Point
Charges}, Ann. Phys., {\bf 301}, 85-127 (2002).

\bibitem[NP1]{NP1} D. Noja, A. Posilicano, \emph{The Wave Equation with
One Point Interaction and the (Linearized) Classical Electrodynamics
of a Point Particle}, Ann. Inst. Henri Poincar\'e, {\bf 68}, 351-377 (1998).

\bibitem[NP2]{NP2} D. Noja, A. Posilicano, \emph{On the Point Limit of
the Pauli-Fierz Model}, Ann. Inst. Henri Poincar\'e, {\bf 71}, 425-457 (1999).

\bibitem[P1]{P1} A. Posilicano, \emph{A Kre\u\i n-like Formula for Singular
Perturbations of Self-Adjoint Operators and Applications},
J. Funct. Anal., {\bf 183}, 109-147 (2001).

\bibitem[P2]{P2} A. Posilicano, \emph{Singular
Perturbations of Abstract Wave Equations},
J. Funct. Anal. (at press)

\bibitem[RS1]{RS1} M. Reed, B. Simon, \emph{Methods of Modern
Mathematical Physics. Vol 1: Functional Analysis}, Academic Press (1972).

\bibitem[RS2]{RS2} M. Reed, B. Simon, \emph{Methods of Modern
Mathematical Physics. Vol 2: Fourier Analysis, Self-Adjointness}, Academic Press (1975).

\bibitem[RS3]{RS3} M. Reed, B. Simon, \emph{Methods of Modern
Mathematical Physics. Vol 3: Scattering Theory}, Academic Press (1979).

\bibitem[RS4]{RS4}M. Reed, B. Simon, \emph{Methods of Modern
Mathematical Physics. Vol 4: Analysis of Operators}, Academic Press (1978).

\bibitem[SW1]{SW1} A. Soffer, M.I. Weinstein, \emph{Nonautonomous
Hamiltonians}, J. Stat. Phys., {\bf 93}, No. 1-2, 359-391
(1998).

\bibitem[SW2]{SW2} A. Soffer, M.I. Weinstein, \emph{Time Dependent
Resonances Theory}, Geom. Funct. Anal., {\bf 8}, 1-43 (1998).

\bibitem[SW3]{SW3} A. Soffer, M.I. Weinstein, \emph{Resonances,
Radiation Damping and Instability in Hamiltonian Nonlinear Wave
Equations}, Invent. Math., {\bf 136}, 9-74 (1999).

\bibitem[S]{Spohn} H. Spohn, \emph{Dynamics of Charged Particles and their Radiation Field},
Cambridge Univ. Press (2004).

\bibitem[T]{T} J.D. Templin, \emph{Radiation Reaction and Runaway Solutions in Acoustics},
Am. J. Phys., {\bf 67}, No. 5, 407-413 (1999).

\end{thebibliography}
\end{document}